\documentstyle[11pt]{article}
\renewcommand{\title}[1]{%
    \bigskip%
    \begin{center}%
    \Large\bf #1%
    \end{center}%
    \vskip .2in}

\renewcommand{\author}[1]{%
    {\begin{center}
    #1
    \end{center}}}
\newcommand{\address}[1]{\vspace{-1.7em}\vspace{0pt}
    {\begin{center}
    \it #1
    \end{center}}}
\begin{document}
\begin{titlepage}
\title{Schwinger Model In Noncommutating Space-Time}

\author{Anirban Saha $\,^{\rm a,b}$,
Anisur Rahaman $\,^{\rm c,d}$,
Pradip Mukherjee\footnote{Also Visiting Associate, S. N. Bose National Centre 
for Basic Sciences, JD Block, Sector III, Salt Lake City, Calcutta -700 098, India and\\ IUCAA, Post Bag 4, Pune University Campus, Ganeshkhind, Pune 411 007,India }
$\,^{\rm a,e}$\ }
\address{$^{\rm a}$Department of Physics, Presidency College\\
86/1 College Street, Kolkata - 700 073, India}
\address{$^{\rm c}$Department of Physics, Durgapur Govt. College,\\
Durgapur - 713214, Burdwan, West Bengal, India}
\address{$^{\rm b}$\tt ani\_saha09@dataone.in}
\address{$^{\rm d}$\tt anisur.rahman@saha.ac.in}
\address{$^{\rm e}$\tt pradip@bose.res.in}

\begin{abstract}
The $1+1$ dimensional bosonised Schwinger model has been studied in a noncommutative scenario. The theory in the reduced phase space exhibits a massive boson interacting with a background. The emergence of this background interaction is a novel feature due to noncommutativity. The structure of the theory ensures unitarity and causality.
\end{abstract}

\noindent {\bf PAC codes:} 11.15.-q, 11.10.Nx,  

\end{titlepage}

The idea of fuzzy space time where the coordinates $x^{\mu}$ satisfy the
noncommutative (NC) algebra 
\begin{equation}
\left[x^{\mu}, x^{\nu}\right] = i \theta^{\mu \nu}
\label{ncgeometry}
\end{equation}
with constant anti-symmetric $\theta^{\mu \nu}$, was mooted
long ago \cite{sny}. This idea has been revived in the recent past and
field theories defined over this NC space has been studied extensively \cite{szabo}.
A contentious issue in this scenario is the aspect of noncommutativity in the time-space sector. It was argued that introduction of space-time noncommutativity spoils unitarity \cite{gomis, gaume} or even causality \cite{sei}. However, much attention has been devoted in recent times to circumvent these difficulties in formulating theories with $\theta^{0i} \neq 0$ \cite{bala, bala1, Daiy}. The $1+1$ dimensional field theoretic models are particularly important in this context because any noncommutative extension of such models essentially contains fuzziness in the time-space sector. The two dimensional field theories have long been recognized as the laboratory where important ideas can be tested in a simple setting. However, not much analysis of the corresponding NC theories is available in the literature. In the present paper we will consider the $1+1$ dimensional Schwinger model \cite{SCH} on a NC setting. Apart from the specific NC aspect such studies are also motivated by the inherent interest of the model as $1+1$ dimensional electrodynamics. 

Historically, the first two dimensional model was proposed  by Thirring \cite{TH}
describing a pure fermionic current-current interaction. The interest increased a huge when Schwinger was able to obtain an exact solution of two dimensional electrodynamics with massless spinor \cite{SCH}. The Schwinger model i.e., the theory of mass less fermion interacting with
an Abelian gauge field in $(1+1)$ dimensional space time is an exactly solvable
field theoretical model. It has been extensively studied over the years \cite{old1, old2, old3, old4,  old5, JR, rb, GB, PM, ARP, kij, AR, dur, sad, san, siy} mainly due to the emergence of phenomena such as mass generation and confinement of fermions (quarks). The gauge field acquires a mass via a kind of dynamical symmetry breaking and the fermions disappear from the physical spectra. 

In $1+1$ dimensions an exact mapping can be established between the bosonic and fermionic theories. The singularities of the Schwinger model can be accommodated by regularizing the fermionic current. An equivalent approach is to obtain the effective action by integrating out the fermions. Commonly it is known as bosonization of the
Schwinger model. For extension to NC scenario the bosonised version is more economical.
So we will follow this approach in the present paper. 

The Schwinger model is defined by the Lagrangian density
\begin{equation}
{\cal L}_F = \bar\psi (i\partial\!\!\!\!/ -eA\!\!\!\!/)\psi
-{1\over 4}F^{\mu\nu}F_{\mu\nu}, 
\label{1} 
\end{equation}
where   the  Lorentz  indices  run  over  the  two  values  $0,1$ and the rest of  the
notation  is  standard. Notice that the coupling constant $e$ has
unit mass dimension in this  situation. Bosonizing the fermion field we get the Lagrangian density involving a scalar field $\phi$ instead of the Dirac field $\psi$:
\begin{equation}
{\cal  L}_B ={1\over 2}\partial_\mu\phi\partial^\mu\phi
+\frac{e}{2}\epsilon_{\mu\nu}F^{\mu\nu}\phi
+ae^2 A_\mu A^\mu
 +{\alpha\over 4}F_{\mu\nu}F^{\mu\nu}.\label{2}
\end{equation}
The first piece is the kinetic energy term for the  scalar  field
and the second one describes the interaction between the matter field and
gauge field. The last two terms involve something new, {\it  viz.}, two undetermined parameters
$a$ and $\alpha$.\footnote{Note that the usual kinetic energy term is absorbed within $\alpha$.} These are fallouts of the regularization process.
To be more specific, if the left handed and the right handed component of $\psi$  are  integrated  out  one by one the  regularization of the determinant contains such parameters \cite{JR, PM, AR}. Setting $a$ to be zero and $\alpha$ to be $-1$ we get the bosonized version of the usual gauge invariant vector Schwinger model. We will use this version for the NC extension in the sequel. 

 The NC action which we consider is 
\begin{equation}
\int d^2x {\cal  L}_{NCBS} = \int d^2x \left[{1\over 2}\left(\hat{D}_\mu\star \hat\phi\right)
\star \left(\hat{D}^\mu\star\hat{\phi} \right)
+\frac{1}{2}e\epsilon^{\mu\nu}\hat{\phi}\star\hat{F}_{\mu\nu}
 - {1\over 4}\hat{F}_{\mu\nu} \star \hat{F}^{\mu\nu}\right].
\label{NCBS}
\end{equation}
where $\hat \phi$ is the NC scalar field and $\hat A_{\mu}$ is the $\star$-gauge
field. We adopt the Minkowski metric $\eta_{\mu \nu} = {\rm diag} \left( +,-\right)$. The covariant derivative $\hat D_{\mu}\star \hat \phi$ is defined as
\begin{equation}
\hat D_{\mu}\star \hat \phi = \partial_{\mu} \hat \phi - i \left[\hat A_{\mu}, \phi\right]_{\star}
\label{covder}
\end{equation}
Here $\star$ denotes that the ordinary multiplication is replaced by the star multiplication defined by
\begin{equation}
\hat \phi(x) \star \hat \psi(x) = \left(\hat \phi \star \hat \psi \right)(x) = e^{\frac{i}{2}
\theta^{\alpha\beta}\partial_{\alpha}\partial^{'}_{\beta}} 
  \hat \phi (x) \hat \psi(x^{'})\big{|}_{x^{'}=x.} 
\label{star}
\end{equation}
The action (\ref{NCBS}) is invariant under the $\star$-gauge
transformation
\begin{equation}
\hat \delta_{\hat \lambda} \hat A_{\mu} = \hat D_{\mu} \star \hat \lambda, \qquad \hat \delta_{\hat \lambda} \hat \phi = -i \left[\hat \phi, \hat \lambda  \right]_{\star}
\label{starg}
\end{equation}

The physics behind the NC theory (\ref{NCBS}) can be explored by several approaches which sometimes compliment each other \cite{pmas}. Thus one can think of the fields as operators carrying the realization of the basic algebra (\ref{ncgeometry}) or a conventional phase space may be used where the ordinary product is deformed. A particularly interesting scenario appears in case of the gauge theories where one can use the Seiberg--Witten (SW) type transformations \cite{SW, bichl, vic} to construct commutative equivalent models \cite{all1, all2, all3, all4} of the actual NC theories in a perturbative framework. In the present paper we adopt this approach. Note that even if the fields and the coordinates in the commutative equivalent model are commuting it is not obvious that the usual Hamiltonian procedure could produce dynamics with respect to noncommuting time. This issue has been addressed by Dayi \cite{Daiy} where noncommutativity in time space sector emerges in a theory with spatial noncommutativity due to a duality transformation. Specifically a Hamiltonian formulation was obtained with commutating time which was shown to be identical to order $\theta$ for both the original theory ( with noncommutativity in the spatial sector only ) and its dual containing space time noncommutativity \cite{Daiy}. Following this we propose to carry out our analysis to first order in $\theta$ and assume the applicability of the usual Hamiltonian dynamics for the commutative equivalent model. 
Since our motivation is to investigate what new features emerge from the presence of noncommutativity in the Schwinger model, introduction of minimal noncommutativity will be sufficient.  

To the lowest order in $\theta$ the explicit forms of the SW maps are known as \cite{SW,bichl, vic}
\begin{eqnarray}
\hat \phi &=& \phi - \theta^{mj}A_{m}\partial_{j}\phi \nonumber\\
\hat A_{i} &=& A_{i} - \frac{1}{2}\theta^{mj}A_{m}
\left(\partial_{j}A_{i} + F_{ji}\right)
\label{1stordmp}
\end{eqnarray}
Using these expressions and the star product (\ref{star}) to order $\theta$ in (\ref{NCBS}) we get
\begin{eqnarray}
\hat S &\stackrel{\rm{SW \; map}}{=}& \int d^{2}x \left[\left\{ 1 + \frac{1}{2} {\rm{Tr}}\left(F \theta\right)\right\}{\mathcal{L}}_{c} - \left(F \theta\right)_{\mu}{}^{\beta}\partial_{\beta}\phi \partial^{\mu}\phi - \frac{e}{2} \epsilon^{\mu \nu}\left(F \theta F\right)_{\mu \nu} \phi \right.\nonumber\\
&& \qquad\qquad \left. + \frac{1}{2}\left(F \theta F\right)^{\mu \nu}F_{\mu \nu}\right]
\label{1storderac}
\end{eqnarray}
where ${\mathcal {L}}_{c}$ stands for the commutative Lagrangean
\begin{eqnarray}
{\mathcal {L}}_{c} =  \frac{1}{2}\partial_{\mu}\phi \partial^{\mu}\phi + \frac{e}{2}\epsilon_{\mu\nu}F^{\mu\nu}\phi - {1\over 4}F_{\mu\nu}F^{\mu\nu} 
\label{cL}
\end{eqnarray}
Note that we can write the action (\ref{1storderac}) in a form which, modulo total derivative terms, does not contain second or higher order time derivatives. This happens because we are considering a perturbative calculation to first order in $\theta$. 
If we would calculate to second order or beyond, higher derivatives of time would appear in the Lagrangean from star product expansion which brings complication in the Hamiltonian formulation \cite{Git, EW}. 

 The Eular--Lagrange equations following from the action (\ref{1storderac}) are
\begin{eqnarray}
&&\partial_{\xi}\left[\left\{ 1 + \frac{1}{2} {\rm{Tr}}\left(F \theta\right)\right\}\partial^{\xi} \phi - \left(F \theta + \theta F\right)^{\xi\mu}\partial_{\mu}\phi \right] - \frac{e}{2}\epsilon^{\mu \nu}\left[\left\{ 1 + \frac{1}{2} {\rm{Tr}}\left(F \theta\right)\right\}F_{\mu\nu} + \left(F \theta F\right)_{\mu \nu}\right] = 0\nonumber\\
&&\partial_{\xi}\left[- \theta^{\xi\alpha}{\mathcal {L}}_{c} + \left\{ 1 + \frac{1}{2} {\rm{Tr}}\left(F \theta\right)\right\}\left(e \epsilon^{\xi\alpha}\phi - F^{\xi\alpha}\right) - \theta^{\alpha\mu}\partial_{\mu}\phi\partial^{\xi}\phi + \theta^{\xi\mu}\partial_{\mu}\phi\partial^{\alpha}\phi \right. \nonumber\\
&&\qquad\quad \left.
- e\phi \left\{\epsilon^{\xi\mu}\left(\theta F\right)^{\alpha}{}_{\mu} - \epsilon^{\alpha\mu}\left(\theta F\right)^{\xi}{}_{\mu} \right\} + \left(F F \theta + \theta F F\right)^{\xi \alpha} + \left(F \theta F\right)^{\xi \alpha}
\right] = 0
\label{eqm} 
\end{eqnarray}
We work out the canonical momenta conjugate to $\phi$ and $A_{\alpha}$ respectively:
\begin{eqnarray}
\pi_{\phi} &=& \left\{ 1 + \frac{1}{2} {\rm{Tr}}\left(F \theta\right) - \left(F \theta + \theta F\right)^{00}\right\}\dot\phi - \left(F \theta + \theta F\right)^{0i}\partial_{i}\phi \nonumber\\
\pi^{\alpha} &=& - \theta^{0\alpha}{\mathcal {L}}_{c} + \left\{ 1 + \frac{1}{2} {\rm{Tr}}\left(F \theta\right)\right\}\left(e \epsilon^{0\alpha}\phi - F^{0\alpha}\right) - \theta^{\alpha\mu}\partial_{\mu}\phi\partial^{0}\phi \nonumber\\
&& \qquad\qquad + \theta^{0\mu}\partial_{\mu}\phi\partial^{\alpha}\phi - e\phi \left[\epsilon^{0\mu}\left(\theta F\right)^{\alpha}{}_{\mu} - \epsilon^{\alpha\mu}\left(\theta F\right)^{0}{}_{\mu} \right] \nonumber\\
&& \qquad\qquad + \left(F F \theta + \theta F F\right)^{0 \mu} + \left(F \theta F\right)^{0 \mu}
\label{momenta}
\end{eqnarray}
From (\ref{momenta}) we get after a few steps  
\begin{eqnarray}
\pi_{\phi} &=& \dot{\phi} + \theta F_{01}\dot{\phi}\nonumber\\
\pi^{0} &=& 0 \nonumber\\
\pi^{1} &=& F_{01} +e\phi + \frac{\theta}{2}\left[{\dot{\phi}}^{2} - \left(\partial_{1}\phi\right)^{2} + 3 F_{01}^{2} \right]
\label{1stmomenta}
\end{eqnarray}
 The Hamiltonian follows as 
\begin{eqnarray}
\int dx {\cal H}_{CEV} & = &\int dx \left[{\cal H}_{CS} + \frac{\theta}{2}\left\{ 3\pi_{\phi}^{2}\pi^{1} + 3 e \phi \phi^{\prime2} - 3\pi^{1} \phi^{\prime 2} -3 e \phi \pi_{\phi}^{2}\right.\right.\nonumber\\
&&\qquad\qquad\quad\left.\left.  + 3 e \phi \left(\pi^{1}\right)^{2}- \left(\pi^{1}\right)^3 - 3 e^{2}\phi^{2} \pi^{1} + e^{3}\phi^{3}\right\}\right]
\end{eqnarray}
where ${\cal H}_{CS}$ is given by 
\begin{equation}
{\cal  H}_{CS}=  {1\over  2}\left[\pi_\phi^2+ (\pi^{1}){}^{2} + \phi'^2 + e^2\phi^2\right]
+\pi^1A_0'- e\pi^{1}\phi
\label{hcs}
\end{equation}
Note that this $\theta$-independent term is nothing but the Hamiltonian of the commutative theory.

From the second equation of (\ref{momenta}) we get one primary constraint 
\begin{equation}
\pi^{0} \approx 0
\label{1stc}
\end{equation}
Conserving this in time a secondary constraint
\begin{eqnarray}
\partial_{1}\pi^{1} \approx  0
\label{2ndc}
\end{eqnarray}
 emerges. The constraints (\ref{1stc}, \ref{2ndc}) have vanishing Poission brackets with the Hamiltonian as well as between themselves. No new constraint, therefore, is obtained. It is interesting to note that the constraint structure is identical with the commutative Schwinger model. This structural similarity  is remarkable because the gauge field in our commutative equivalent theory is the SW map of a NC gauge field belonging to the Groenewold--Moyal deformed $C^{\star}$ algebra.

 To proceed further we require to eliminate the gauge redundancy in the equations of motion (\ref{eqm}) by invoking appropriate gauge fixing conditions. The structure of the constraints (\ref{1stc}, \ref{2ndc}) suggests the choices : $A_{0} = 0$ and $A_{1} = 0$. The simplectic structure in the reduced phase space will be obtained from the Dirac brackets which enables to impose the constraints strongly \cite{dir}. It is easy to verify that the nontrivial Dirac brackets of the reduced phase space remains identical with the corresponding canonical brackets. The hamiltoinan density in the constrained subspace can be written down as
\begin{eqnarray}
{\cal H}_R = \frac{1}{2}\left[(\pi_\phi^2+\phi'^{2} + e^{2}\phi^{2})
+ e\theta\phi \left(3\phi'^{2} - 3\pi_{\phi}^{2} + e^{2}\phi^{2}\right)\right] 
\label{HR}
\end{eqnarray}

The Reduced Hamiltonian (\ref{HR}) along with the Dirac brackets
leads to the following equations of motion.
\begin{equation}
\dot\phi=\left(1- 3 e\theta\phi\right) \pi_{\phi} 
\label{EQ1}
\end{equation}
\begin{equation}
\dot\pi_{\phi} = \phi'' - e^{2} \phi + \frac{1}{2} e \theta \left( 3\pi_{\phi}^{2} - 3\phi'^{2} -3 e^{2} \phi^{2}\right)
+ 3 e\theta\left(\phi \phi'' + \phi'^{2}\right)
\label{EQ2}
\end{equation}
The equation (\ref{EQ1}) and (\ref{EQ2}) after a little algebra reduced to
\begin{equation}
(\Box+e^2)\phi =  \frac{3}{2} e \theta \left(-\dot{\phi}^{2} + \phi'^2 + e^{2}\phi^{2} \right)
\label{box}
\end{equation}
Equation (\ref{box}) is the relevant equation of motion obtained by removing the gauge arbitrariness of the theory. To zero order in $\theta$ it gives  
\begin{equation}
(\Box+e^2)\phi=0
\label{spec}
\end{equation}
The equation (\ref{box}) looks complicated but, thanks to the fact that our theory is order-$\theta$, in the right hand side we can substitute $\phi$ to $0$-order. Again, to this order $\phi$ satisfies (\ref{spec}) the solution to which can easily be expanded in terms of the plane wave solutions
\begin{equation}
\phi= \int \frac{d\bar{p}}{\left(2\pi\right)}\frac{1}{\sqrt{2 p_{0}}}\left[a(\bar{p})e^{-ip^{0}x^{0} + i\bar{p}\bar{x}} + a^{\dag}(\bar{p})e^{ip^{0}x^{0}- i\bar{p}\bar{x}} \right] \\
\label{planewave}
\end{equation}
where  $x^{\mu} \equiv \left(x^{0}, \bar{x} \right)$, $p^{\mu} \equiv \left(p^{0}, \bar{p} \right)$ and $p_{0} = \sqrt{\bar{p}^{2} + e^{2}}$. Hence from (\ref{box}) we get 
\begin{equation}
(\Box+e^2)\phi= j(x)
\label{boxf}
\end{equation}
where $j(x)$ is given by 
\begin{eqnarray}
j(x)& = &\frac{3 e \theta}{2} \int \frac{d\bar{p} d\bar{q}}{\left(8\pi^2\right)}\frac{e^{i\left(\bar{p}+\bar{q}\right)\bar{x}} }{\sqrt{p^{0}q^{0}}} \left[\left(p_{0} q_{0} - \bar{p}\bar{q} + e^{2}\right) \left\{a(\bar{p})a(\bar{q})e^{- i\left(p^{0}+q^{0}\right) x^{0}} +
a^{\dagger}(-\bar{p})a^{\dagger}(-\bar{q})e^{ i\left(p^{0}+q^{0}\right) x^{0}} \right\}\right. \nonumber\\
&&\left.+ 
\left( - p_{0} q_{0} - \bar{p}\bar{q} + e^{2}\right) \left\{a(\bar{p})a^{\dag}(- \bar{q})e^{- i\left(p^{0} - q^{0}\right) x^{0}} + a^{\dagger}(-\bar{p})a(\bar{q})e^{ i\left(p^{0} - q^{0}\right) x^{0}}\right\}\right]
\label{j}
\end{eqnarray}
We thus have a bosonic field $\phi$ interacting with a source $j(x)$. It is easy to recognise that the equations (\ref{boxf}) represents the Klien--Gordon (KG) theory with a classical source.

A remarkable observation is inherent in (\ref{boxf}). To understand the proper perspective we have to briefly review the results from the corresponding commutative theory. There we end up with (\ref{spec}) 
and interpreat that the photon has acquired mass and the fermion has disappeared from the physical spectrum \cite{SCH, old1, old2, old3,old4, old5}. This means the fermion is confined. The introduction of noncommutativity changes the scenario in a fundamental way. The gauge boson again acquires mass but this time it is interacting with a background. The origin of this background interaction is the fuzziness of spacetime. This is a physical effect carrying NC signature. In this context it may be mentioned that generation of interactions by casting non-interacting theories in NC coordinates have been observed in other contexts also \cite {ref1, ref2, ref3}.

It is easy to formulate the theory guided by (\ref{boxf}) as a quantum theory. Note that the NC parameter $\theta$ is a small number and we can treat the interaction term as a perturbation. The resulting $S$-matrix can easily be written down as 
\begin{eqnarray}
S \sim T\left\{\exp\left[ -i \int d^{2}x j(x)\phi(x)\right]\right\}
\label{s}
\end{eqnarray}
Since $j(x)$ is real $S$ is unitary.  Again, since the field $\phi$ satisfies a KG equation causality is also ensured.
 These are good news in view of the presence of time-space noncommutativity and justifies our proposition based on \cite{Daiy} that usual Hamiltonian analysis is applicable in our commutative equivalent model.
 
 In this letter we have studied the effect of time-space noncommutativity on the bosonised Schwinger model in $1+1$ dimensions. We analyzed the model in the commutative equivalent representation \cite{all1,all2, all3, all4, pmas} using a perturbative Seiberg--Witten map \cite{SW, bichl, vic}. Following \cite{Daiy} we have assumed the applicability of a usual Hamiltonian analysis of the commutative equivalent model. the  The model exhibits emergence of a massive boson. This is similar to what happens in the usual theory. However, the boson is no longer free as in the commutative counterpart but it interacts with a background. Our analysis thus reveals the presence of a background interaction term which is manifest only in the tiny length scale $\sim \sqrt \theta$. The theory in the reduced phase space can be formulated as a perturbative quantum field theory which is formally similar to the KG theory with a classical source. Consequently, the requirements of unitarity and causality are satisfied. 
\section*{Errata}
The expression of the Hamiltonian (equation 14) contains calculational error. The correct form will be 
\begin{eqnarray}
{\cal H}_{CEV} & = &
\left[{\cal H}_{CS} + \frac{\theta}{2}
\left\{\pi^{1}\left(\phi^{\prime}{}^{2} - \pi_{\phi}{}^{2}\right) + e \phi \left(\pi_{\phi}{}^{2} - \phi^{\prime}{}^{2}\right)\right.\right.\nonumber\\
&&\left. \left.
+ 
e^{3} \phi^{3} - \left(\pi^{1}\right)^{3} + \frac{3}{2} e \phi \pi^{1}\left(\pi^{1} - e\phi\right)\right\}\right]
\label{EncH}
\end{eqnarray}
The error in the Hamiltonian percolates to the subsequent equations (18), (19), (20), (21) and (25) the correct forms of which can be derived using the corrected Hamiltonian (\ref{EncH}). Thus the corrected form of the reduced Hamiltonian is
\begin{eqnarray}
{\cal H}_{R} & = &
\left[ \frac{1}{2} \left(\pi_\phi^2  + \phi^{\prime}{}^{2} + \phi^{2}\right) + \frac{e\theta}{2}\phi\left(\pi_{\phi}^{2} -  \phi^{\prime 2} + e^{2}\phi^{2}\right)\right]
\label{EncHr}
\end{eqnarray}
and the equations of motion for $\phi$ and $\pi_{\phi}$ are 
\begin{eqnarray}
\dot\phi & = & \left(1 + e \theta \phi\right) \pi_{\phi} 
\label{EEQ1}
\end{eqnarray}
\begin{eqnarray}
\dot\pi_{\phi} & = & \phi^{\prime \prime} - e^{2}\phi - \frac{e\theta}{2} \left(\pi_{\phi}^{2} + \phi^{\prime}{}^{2} + 2 \phi \phi^{\prime \prime} + 3 e^{2}\phi^{2}\right) 
\label{EEQ2}
\end{eqnarray}
respectively.
These two equations combine to give the second order equation that contains the physical contents of the theory
\begin{eqnarray}
\left( \Box + e^{2} \right)\phi & = & \frac{e\theta }{2} \left(\dot{\phi}^{2} - \phi^{\prime}{}^{2} - 5 e^{2}\phi^{2} \right) 
\label{Ebox}
\end{eqnarray}
The expression for the source of the interacting background of noncommutative origin is 
\begin{eqnarray}
j(x)& = &\frac{e \theta}{2} \int \frac{d\bar{p} d\bar{q}}{\left(8\pi^2\right)}\frac{e^{i\left(\bar{p}+\bar{q}\right)\bar{x}} }{\sqrt{p^{0}q^{0}}} \left[ \left( - p_{0} q_{0} + \bar{p}\bar{q} - 5 e^{2}\right)\right. \nonumber\\ 
&& \left. \times\left\{a(\bar{p})a(\bar{q})e^{- i\left(p^{0}+q^{0}\right) x^{0}} \right. \right.\nonumber\\
&&\left.\left.+ 
a^{\dagger}(-\bar{p})a^{\dagger}(-\bar{q})e^{ i\left(p^{0}+q^{0}\right) x^{0}} \right\}\right. \nonumber\\
&&\left.+ 
\left( p_{0} q_{0} - \bar{p}\bar{q} + 5e^{2}\right) \left\{a(\bar{p})a^{\dag}(- \bar{q})e^{- i\left(p^{0} - q^{0}\right) x^{0}}\right. \right. \nonumber\\ 
&& \left.\left.+ a^{\dagger}(-\bar{p})a(\bar{q})e^{ i\left(p^{0} - q^{0}\right) x^{0}}\right\}\right]
\label{Ej}
\end{eqnarray}
The equations (\ref{EncHr}), (\ref{EEQ1}), (\ref{EEQ2}), (\ref{Ebox}) and (\ref{Ej}) are the corrected forms of the equations (18), (19), (20), (21) and (25) respectively.
It is easy to appriciate that the corrections do not affect our conclusions in any essential way.
\section* {Acknowledgment}
{AS wants to thank the Council of Scientific and Industrial Research (CSIR), Govt. of India, for financial support and the Director, S. N. Bose National Centre for Basic Sciences for providing computer facilities. He also likes to acknowledge the hospitality of IUCAA where part of this work has been done. Finally the authors thank the referre for his useful comments.}

\end{document}